\documentclass[10pt]{article}

\usepackage{fullpage}
\usepackage{setspace}
\usepackage{parskip}
\usepackage{titlesec}
\usepackage[section]{placeins}
\usepackage{xcolor}
\usepackage{breakcites}
\usepackage{lineno}
\usepackage{hyphenat}
\usepackage{float}
\usepackage{dblfloatfix}
\usepackage{bibunits}

\PassOptionsToPackage{hyphens}{url}
\usepackage[colorlinks = true,
            linkcolor = blue,
            urlcolor  = blue,
            citecolor = blue,
            anchorcolor = blue]{hyperref}
\usepackage{etoolbox}

\usepackage{natbib}

\renewenvironment{abstract}
  {{\bfseries\noindent{\abstractname}\par\nobreak}\footnotesize}
  {\bigskip}

\titlespacing{\section}{0pt}{*3}{*1}
\titlespacing{\subsection}{0pt}{*2}{*0.5}
\titlespacing{\subsubsection}{0pt}{*1.5}{0pt}

\usepackage{authblk}

\usepackage{graphicx}
\usepackage[space]{grffile}
\usepackage{latexsym}
\usepackage{textcomp}
\usepackage{longtable}
\usepackage{tabulary}
\usepackage{booktabs,array,multirow}
\usepackage{amsfonts,amsmath,amssymb}
\providecommand\citet{\cite}
\providecommand\citep{\cite}

\newif\iflatexml\latexmlfalse

\AtBeginDocument{\DeclareGraphicsExtensions{.pdf,.PDF,.eps,.EPS,.png,.PNG,.tif,.TIF,.jpg,.JPG,.jpeg,.JPEG}}

\usepackage[utf8]{inputenc}
\usepackage[english]{babel}

\usepackage{float}

\begin{document}

\title{Mie-resonant silicon waveguide for efficient coupling with excitonic emitters in InSe}

\author[1]{Timofey V. Antipov}%
\author[1, 2]{Alexandra D. Gartman}%
\author[1]{Dmitry N. Gulkin}%
\author[3]{Lyubov Kotova}%
\author[3]{Aidar Galimov}%
\author[3]{Bogdan Borodin}%
\author[3]{Ilya Eliseyev}%
\author[1]{Alexander S. Shorokhov}%
\author[3]{Maxim Rakhlin}%
\author[1]{Andrey A. Fedyanin}%
\affil[1]{Faculty of Physics, Lomonosov Moscow State University, Moscow 119991, Russian Federation}%
\affil[2]{Department of Materials Science, Shenzhen MSU-BIT University, Shenzhen 517182, P.R. China}%
\affil[3]{Ioffe Institute, St. Petersburg 194021, Russian Federation}%

\vspace{-1em}

\date{\today}

\begingroup
\let\center\flushleft
\let\endcenter\endflushleft
\maketitle
\endgroup

\selectlanguage{english}
\begin{bibunit}
\begin{abstract}

Enhancement of radiative coupling efficiency between out-of-plane excitonic emitters in an indium selenide (InSe) film and an integrated waveguide formed by silicon (Si) Mie-resonant nanodisks is experimentally studied. Photoluminescence power at the resonant waveguide output is increased by~2.5 times at 950~nm in comparison with the case of a conventional rib waveguide of the same geometrical parameters due to the efficient excitation of Mie-type magnetic dipole resonances in individual nanoparticles. These results show inspiring possibilities for creating new on-chip light emitters for various integrated photonics applications.

\end{abstract}

\sloppy

\section*{Introduction}


Modern photonic technologies can offer various advantages over purely electronic ones, including faster signal processing and low energy consumption based on specially designed micro- and nanoscale devices. Optical chips for various applications such as LiDARs~\textsuperscript{\hyperref[csl:1]{1}}, 
biosensors~\textsuperscript{\hyperref[csl:2]{2}} and optical computers that transmit and process information using photons are under development~\textsuperscript{\hyperref[csl:3]{3}}. 
The main components of such devices, regardless of the specific task being solved, are light sources and waveguides that transmit and redirect optical radiation~\textsuperscript{\hyperref[csl:4]{4}}. Silicon is one of the most commonly used semiconductors in nanophotonics technology for waveguides, with a transmission region corresponding to the infrared region of the spectrum~\textsuperscript{\hyperref[csl:5]{5,}}~\textsuperscript{\hyperref[csl:6]{6}}. However, silicon is not suitable for direct light generation, therefore efficient Si waveguide-compatible emitters operating effectively at these wavelengths are needed. It defines interest of the integrated photonics community to development of a new generation of light sources on a chip compatible with modern semiconductor technologies for increasing photo- and electroluminescence based on various emitter mechanisms such as Er-related light sources, Ge-on-Si lasers, III-V-based Si lasers and etc~\textsuperscript{\hyperref[csl:7]{7}}.

Special attention is paid to excitonic light sources~\textsuperscript{\hyperref[csl:8]{8}}. They offer several advantages, including strong light-matter interaction, unique optical properties, wide spectral range, the potential for quantum devices and all-optical transistors~\textsuperscript{\hyperref[csl:9]{9}}. They are highly efficient, versatile, and enable precise control of light emission properties~\textsuperscript{\hyperref[csl:10]{10}}. Excitonic devices hold potential for quantum computation, while all-optical transistors offer new possibilities for signal processing~\textsuperscript{\hyperref[csl:11]{11}}. One of the most promising sources of excitonic luminescence with high-efficient coupling to waveguides are 2D materials, due to their large exciton binding energy, which can reach hundreds of meV~\textsuperscript{\hyperref[csl:12]{12}}. 2D materials can be easily integrated into other systems or devices~\textsuperscript{\hyperref[csl:13]{13,}}~\textsuperscript{\hyperref[csl:14]{14}}, as well as assembled into heterostructures with the desired functionality~\textsuperscript{\hyperref[csl:15]{15,}}~\textsuperscript{\hyperref[csl:16]{16}}. 

Transition metal dichalcogenides (TMDs) are widely known 2D materials as perfect sources of excitonic photoluminescence~\textsuperscript{\hyperref[csl:12]{12,}}~\textsuperscript{\hyperref[csl:17]{17,}}~\textsuperscript{\hyperref[csl:18]{18,}}~\textsuperscript{\hyperref[csl:19]{19,}}~\textsuperscript{\hyperref[csl:20]{20}}. The dipole moment orientation of excitonic emitters determines their coupling to waveguides. In case of TMDs, dipole moments of bright excitons are randomly oriented in-plane that lowers the efficiency of their radiation coupling to waveguides~\textsuperscript{\hyperref[csl:21]{21}}. Contrary, gray excitons, found in thin films of layered III-VI monochalcogenides like InSe, can be potentially suitable to solve this problem~\textsuperscript{\hyperref[csl:22]{22,}}~\textsuperscript{\hyperref[csl:23]{23,}}~\textsuperscript{\hyperref[csl:24]{24}}. These excitons allow the electron and hole to be separated into different layers, creating a dipole with an out-of-plane orientation~\textsuperscript{\hyperref[csl:25]{25}}. 

InSe offers the large band-gap tunability: its optical band gap blue-shifts as the number of layers decrease~\textsuperscript{\hyperref[csl:26]{26}}. InSe has weak electron-phonon scattering~\textsuperscript{\hyperref[csl:27]{27}} and a low effective mass~\textsuperscript{\hyperref[csl:28]{28,}}~\textsuperscript{\hyperref[csl:29]{29,}}, resulting in high carriers mobility. The absence of inversion symmetry in certain polytypes of InSe crystals enables second-harmonic generation for all layer thicknesses~\textsuperscript{\hyperref[csl:30]{30}}. These properties make InSe a great choice for various photonic and optoelectronic applications~\textsuperscript{\hyperref[csl:31]{31,}}~\textsuperscript{\hyperref[csl:32]{32}}~\textsuperscript{\hyperref[csl:33]{33}}.

Efficient delivery of radiation from the emitter directly to the integrated circuit requires strong optical coupling between the emitter and the on-chip waveguide system. Such on-chip systems can be a compound of a large number of subwavelength particles supporting excitation of optical resonances. All-dielectric nanoparticles made of transparent materials with a high refractive index are becoming increasingly popular alternatives to metal-based ones for light control in an integrated circuit~\textsuperscript{\hyperref[csl:34]{34,}}~\textsuperscript{\hyperref[csl:35]{35,}}~\textsuperscript{\hyperref[csl:36]{36,}}~\textsuperscript{\hyperref[csl:37]{37,}}~\textsuperscript{\hyperref[csl:38]{38}}. These particles support excitation of both electric (ED) and magnetic (MD) dipole Mie resonances and are compatible with modern complementary metal-oxide-semiconductor technologies (CMOS). By arranging such nanoantennas into one-dimensional chains and exciting the MD resonance in each particle, it is possible to achieve efficient propagation of radiation along the chain due to near-field optical coupling between neighboring particles~\textsuperscript{\hyperref[csl:39]{39}}. It has been shown numerically, that placing InSe thin films and TMD heterostructures on such chains leads to increased optical coupling with the waveguide~\textsuperscript{\hyperref[csl:40]{40,}}~\textsuperscript{\hyperref[csl:41]{41}}.

\begin{figure}[H]
\begin{center}
\includegraphics[width=0.6\linewidth]{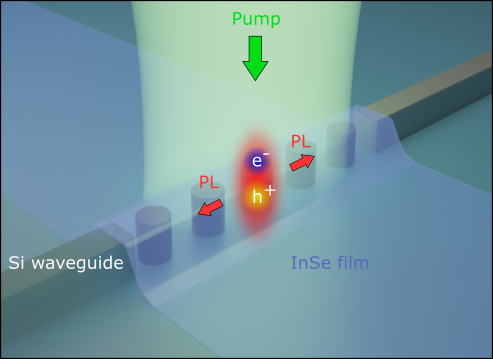} 
\caption{The sketch showing optical coupling between the resonant silicon waveguide structure and the excitonic source localized in the InSe film.}
\label{concept}
\end{center}
\end{figure}

In this study, we experimentally demonstrate efficient coupling between integrated resonant waveguide structure based on Mie-type nanoparticles and thin films of InSe. To this end we fabricate silicon waveguides with embedded Mie nanoantennas and InSe flakes of specific thicknesses placed on the waveguides (see Supplementary Materials, Section~S1 for the fabrication details). They support MD excitation which is field-complementary with excitonic emitters localized on defects in the InSe film covering the waveguide (Fig~{\ref{concept}}). Micro-photoluminescence ($\mu$-PL) experiments demonstrate an increase in emission intensity due to the enhancement of the optical coupling of the emitted light with the waveguide structure on the chip and due to the Purcell effect associated with the local field resonant enhancement. We support our experimental findings with numerical simulations that show the efficient coupling between out-of-plane excitons and integrated waveguides composed of resonant Si nanoparticles.

\par\null\selectlanguage{english}
\par\null

\section*{Sample under study}

\begin{figure}[H]
\begin{center}
\includegraphics[width=0.5\linewidth]{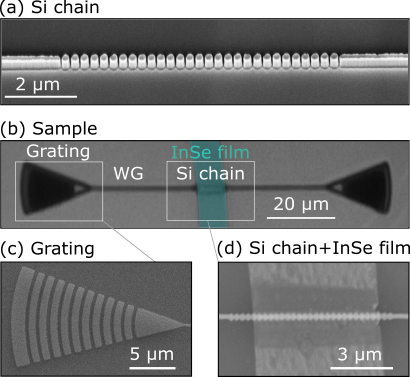} 
\caption{(a) Scanning electron microscope (SEM) image of the chain before the film transfer. (b) Optical microscope image of the sample. (c) SEM image of the grating at the end of the waveguide. (d) SEM image of the chain combined with the InSe film.}
\label{sample}
\end{center}
\end{figure}

A typical array of silicon resonant nanoparticles waveguide (RW) made of a SOI-based wafer with a 2 $\mu$m-thick oxide layer and 220 nm-thick Si layer is shown in Fig~{\ref{sample}}(a--d). The sample consists of several parts: a central part in the form of silicon nanodisks shaped as truncated cones and two 140$\pm$5\,nm-wide strips with the grating couplers at the edges (Fig~{\ref{sample}}(a, b)). The distance between the disks is 50$\pm$5\,nm, the diameter of the disks is 220$\pm$10\,nm and the height of the disks is 170$\pm$5\,nm. Nanoparticle parameters are selected to increase the efficiency of optical coupling between dipole emitters in InSe film and resonant waveguides due to the excitation of the MD resonance in the disks~\textsuperscript{\hyperref[csl:40]{40}}. There are two focusing grating couplers at the ends of the waveguides, which are used to direct light out of the chip for measuring the fraction of radiation coupled to and transmitted through the waveguide (Fig~{\ref{sample}}(c)). A silicon slab with a thickness of 50$\pm$5\,nm is also formed between the substrate and the waveguide system, and the total height of the structure is 220$\pm$10\,nm. Numerical modelling of the system using the finite-difference time-domain (FDTD) method shows that this feature of the manufactured sample does not have any signi\-ficant effect on the optical properties of the system under study. For the $\mu$-PL measurements, a reference array of conventional rib waveguides (CW) without chain of nanodisks is also fabricated.

\par\null\selectlanguage{english}
\par\null

Further, the InSe flake obtained by employing the mechanical exfoliation technique is transferred onto the central part of the waveguide. Fig~{\ref{sample}}(b, c, d) show the combination of the film and RW in the experimental sample. The average film thickness measured by atomic force microscopy is 15$\pm$5\,nm. This value is optimal because, on the one hand, reducing the thickness of the film would lead to a blueshift in the photoluminescence spectrum~\textsuperscript{\hyperref[csl:42]{42}} and, on the other hand, increasing the thickness can negatively affect the transmission of the hybrid waveguide.  

\section*{Optical spectroscopy measurements}

We perform the optical spectroscopy to characterize the resonant waveguide systems fabricated in our expe\-riments. In these measurements, we use a home-built setup, the detailed representation of which is shown in Supplementary Materials, Section~S3.

\par\null\selectlanguage{english}
\begin{figure}[H]
\begin{center}
\includegraphics[width=0.5\linewidth]{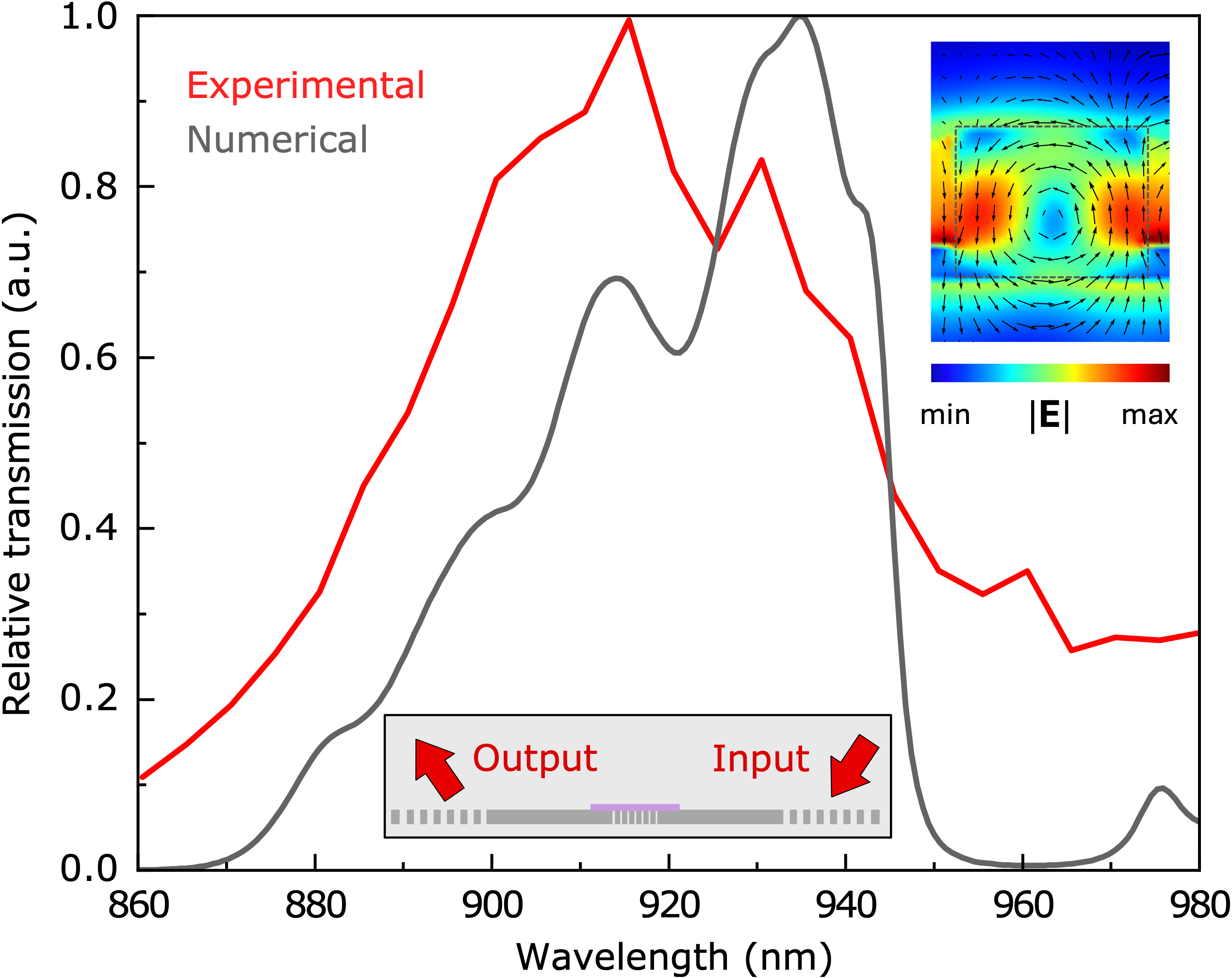} 
\caption{Relative transmission spectra of the waveguides. Inset at the bottom shows a schematic illustration of the experimental configuration. Inset to the right shows the electric field distribution in the central disk cross-section in at 937~nm.}
\label{optical_spectroscopy}
\end{center}
\end{figure}
\par\null

A schematic illustration of the experimental configuration in reflection scheme is shown by the inset image in Fig~{\ref{optical_spectroscopy}}. It enables simultaneous visualization of the entire sample, as well as the focusing of laser light on the input grating. Moreover, it allows the radiation transmitted by the output grating to be collected and spatially filtered. The TM--polarization of the laser beam is selected by the Glan-Taylor prism. The measured transmission spectrum of the RW is further normalized to the same spectrum obtained for the CW. 

The experimental dependence of the relative transmission on the wavelength is shown in Fig.~{\ref{optical_spectroscopy}} by the red curve. The maximum at 915~nm corresponds to the the most efficient optical coupling between the nanoparticles of the chain due to which the radiation propagates along it with the least losses. The relative transmission of the system is numerically calculated using FDTD method. For modelling, we use a Si waveguide 170~nm-high and 140~nm-wide with a central part in the form of a chain of 29~silicon nanodisks with a height of 170~nm and a spacing of 50~nm, covered with a 15~nm-thick InSe film. The disk diameters are set to a uniform distribution in the range 220$\pm$10\,nm in order to take into account fabrication imperfections. As in the experiment, a 50~nm-thick slab is placed between the sample and the oxide layer.

The source of the fundamental TM--mode is placed before the chain of nanodisks, and a power monitor, which is used to measure the transmission, after the chain. A similar simulation is carried out for the rib waveguide of the same length. In the simulations we use the perfectly matched layer boundary conditions. Dividing the spectrum for the nanoparticles chain by the spectrum of the conventional waveguide gives the relative transmission (gray curve in Fig.~{\ref{optical_spectroscopy}}). The blueshift of the experimental spectrum can be explained by the imperfection of the fabricated structure.

\section*{Micro-Raman measurements}

\begin{figure}[H]
\begin{center}
\includegraphics[width=1\linewidth]{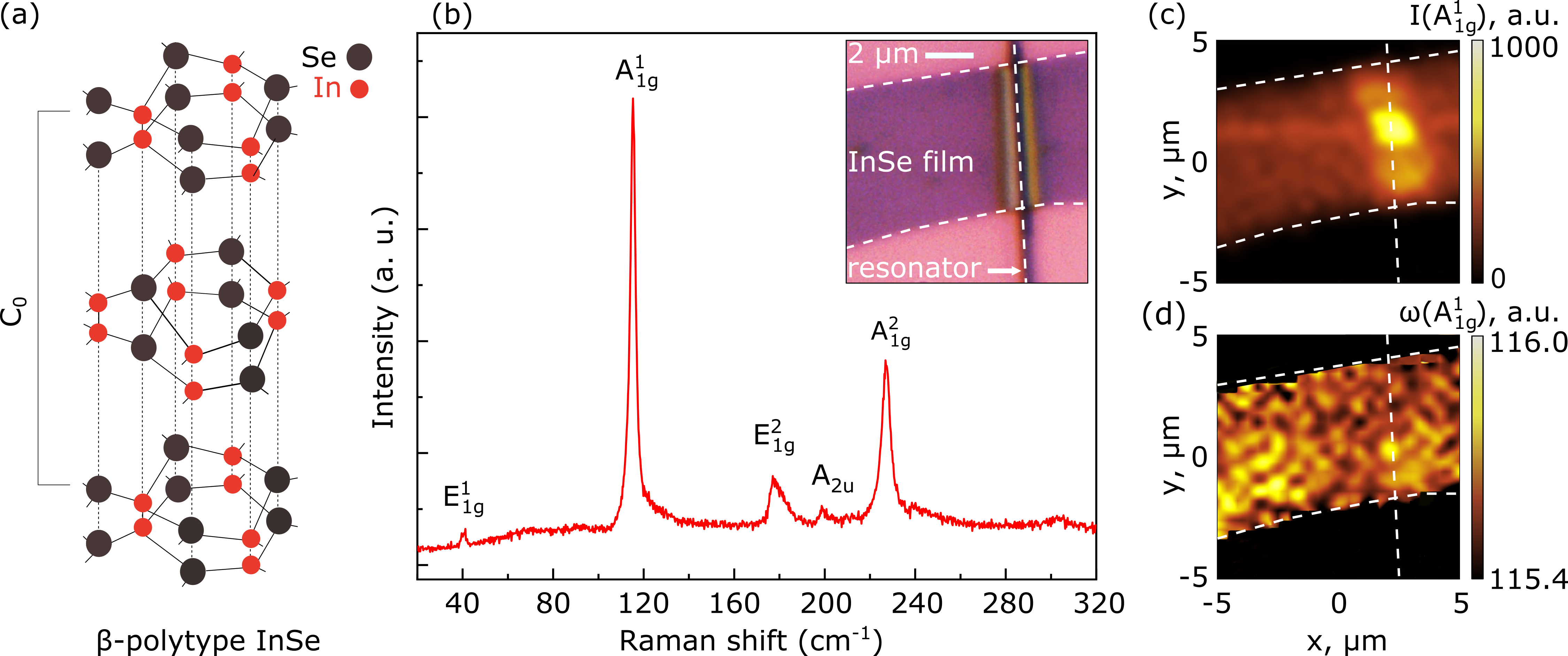} 
\caption{Raman scatering studies: (a)~Unit cell of $\beta$-InSe. (b)~Typical Raman spectrum of the InSe flake under study. Inset shows an optical image of the area of by Raman mapping. (c, d)~Raman scattering  maps demonstrating the peak intensity and frequency distribution of the $A_{1g}^1$ Raman line in the vicinity of the waveguide. Dashed curves are the borders of the flake and the waveguide.}
\label{raman_spectroscopy}
\end{center}
\end{figure}

Raman spectroscopy is employed for structural characterization of the samples. Typical Raman spectrum of the InSe flake (Fig~{\ref{raman_spectroscopy}}(b)) consists of modes characteristic for $\beta-$InSe polytype~\textsuperscript{\hyperref[csl:43]{43}}: E$_\text{g}(1)$ (41~cm$^{-1}$), A$_\text{1g}^{1}$ (115~cm$^{-1}$), E$_\text{g}^{2}$ (178~cm$^{-1}$), E$_\text{u}^{1}$ (186~cm$^{-1}$), A$_\text{1g}^{2}$ (227~cm$^{-1}$) and a weak contribution of A$_\text{2u}$ phonons (198~cm$^{-1}$)~\textsuperscript{\hyperref[csl:44]{44}}. In addition to point measurements, the samples are investigated by Raman mapping. Fig~{\ref{raman_spectroscopy}}(c) represents the intensity distribution of the A$_\text{1g}^{1}$ peak. 
Some enhancement (2-3$\times$) of the Raman signal observing in the flake region above the resonator may be caused by the interference enhancement of the Raman signal by the resonator. The effect is similar to that was observed in thin flakes and films deposited on SiO2/Si substrates~\textsuperscript{\hyperref[csl:45]{45}}. The Raman map of the A$_\text{1g}^{1}$ peak position (Fig~{\ref{raman_spectroscopy}}(d)) reveals that its frequency varied by no more than 0.6~cm$^{-1}$, which is within the spectrometer resolution. According to the data obtained for a 10-15~nm-thick flake~\textsuperscript{\hyperref[csl:46]{46}}, the strain shift rate of the A$_\text{1g}$ Raman line is -2.5~cm$^{-1}$/\%, thus, the variation of strain within the flake is significantly less than~0.24\%. This indicates that, although placing the flake onto a waveguide is expected to cause strain in the InSe flake, the properties of InSe in the vicinity of the resonator are close to the properties of an unstrained InSe flake.

\par\null\selectlanguage{english}
\par\null

\section*{Micro-photoluminescence measurements}

\begin{figure}[H]
\begin{center}
\includegraphics[width=0.9\linewidth]{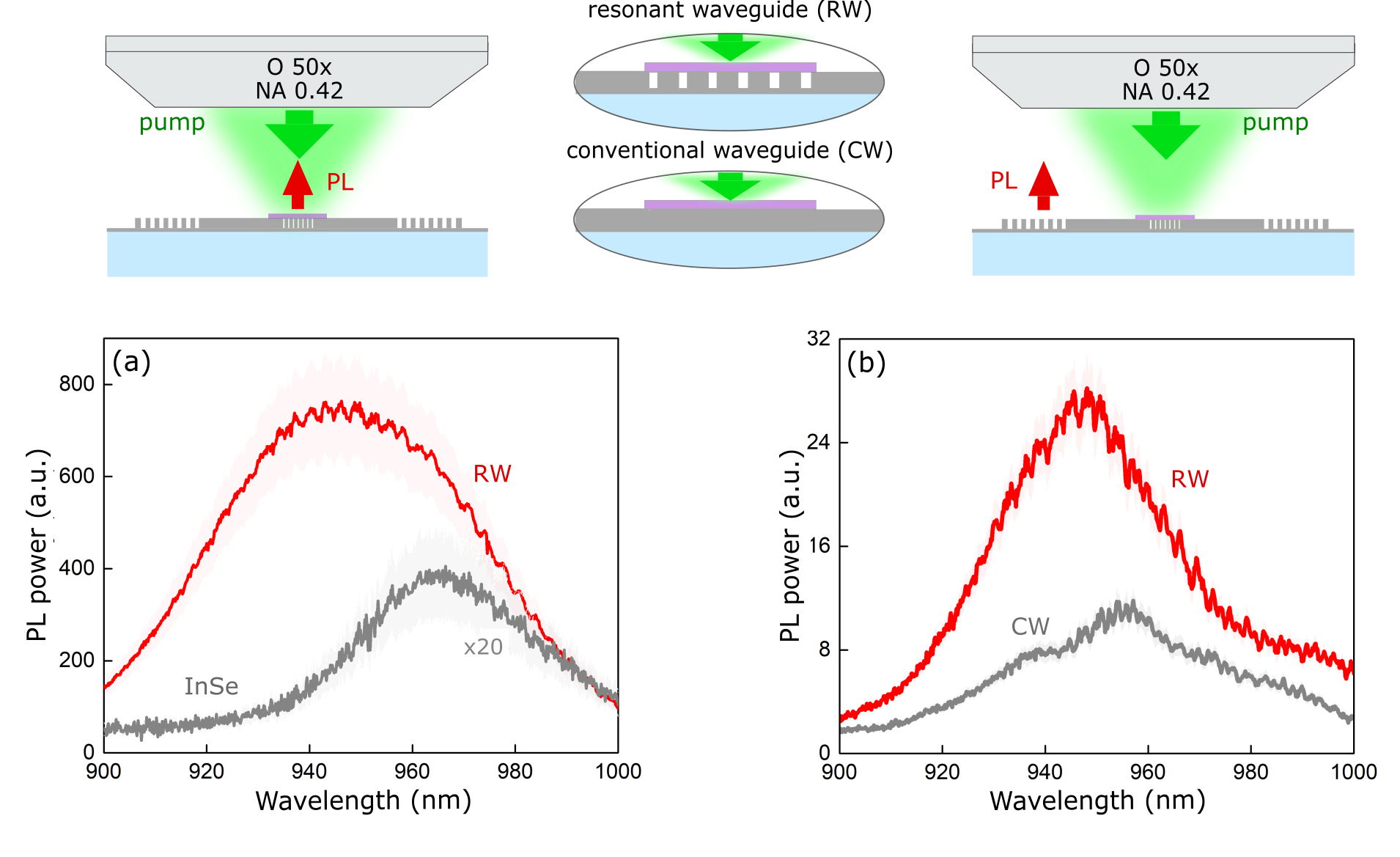} 
\caption{$\mu$-PL measurement results. The sketches above the graphs show the concept of measurements in the corresponding cases. (a)~Spectra of $\mu$-PL emitted in backscattering (red curve corresponds to the RW and gray curve corresponds to the InSe flake on the substrate). (b)~Spectra of $\mu$-PL extracted through the grating coupler (red curve corresponds to the RW and gray curve corresponds to the CW).}
\label{PL_results}
\end{center}
\end{figure}

The $\mu$-PL measurements of the sample are carried out at T~=~8 K using the setup shown in Supplementary Materials, Section~S4. The sketches at the top of the Fig~{\ref{PL_results}}(a, b) illustrate the measurement concepts and the graphs below show the experimental results. The left figure corresponds to the excitation of the film located directly on the RW (red curve) and on the substrate without the structure (gray curve) and the collection of PL radiation emitted in backscattering, while the right figure corresponds to the collection of PL through a grating coupler (radiation from the film in this case is blocked by a diaphragm) of the RW (red curve) and the CW (gray curve). From the Fig~{\ref{PL_results}}(b), one can compare radiation coupling efficiency for the RW and CW. To do this, we divide the PL power of the RW at a wavelength of 950~nm by the corresponding value for CW. As a result, we get a~2.5-fold increase in efficiency.

\par\null\selectlanguage{english}
\par\null

The main factors of PL power enhancement in the experiments are: 
\begin{enumerate} 
\item Local pump field enhancement due to the resonant nanophotonic structure under the InSe film leading to the increased absorption of the pump radiation~\textsuperscript{\hyperref[csl:40]{40}}.
\item Increase in the rate of spontaneous PL emission due to the Purcell effect~\textsuperscript{\hyperref[csl:47]{47}}.
\item Increase in the number of emitters due to local strain in the film~\textsuperscript{\hyperref[csl:21]{21,}}~\textsuperscript{\hyperref[csl:48]{48}}.
\item Optical radiation redirection effect: antennas not only inject radiation into the waveguide, but also re-scatter it into the objective lens. 
\end{enumerate} 

We also calculate the fraction of radiation coupled into the waveguide: 

\begin{displaymath}
  \frac{2 P_{grat}}{2 P_{grat} + P_{back} + P_{loss}}\times100\%,     
\end{displaymath}

\noindent where $P_{grat}$ is the power at one grating (known from the Fig~{\ref{PL_results}}(b)), $P_{back}$ is the power in the back reflection (known from the Fig~{\ref{PL_results}}(a)) and $P_{loss}$ is the loss power (power of radiation absorbed in the waveguide). The last parameter is estimated from the calculated transmission of the waveguide (see Supplementary Materials, Section~S5). According to the Fig~{\ref{PL_results}}, the final result for the efficiency of the radiation coupling into the waveguide is~33$\%$ for the RW and~13$\%$ for the CW. 

\section*{Contributions from variously oriented dipole emitters in InSe}

\begin{figure}[H]
\begin{center}
\includegraphics[width=0.8\linewidth]{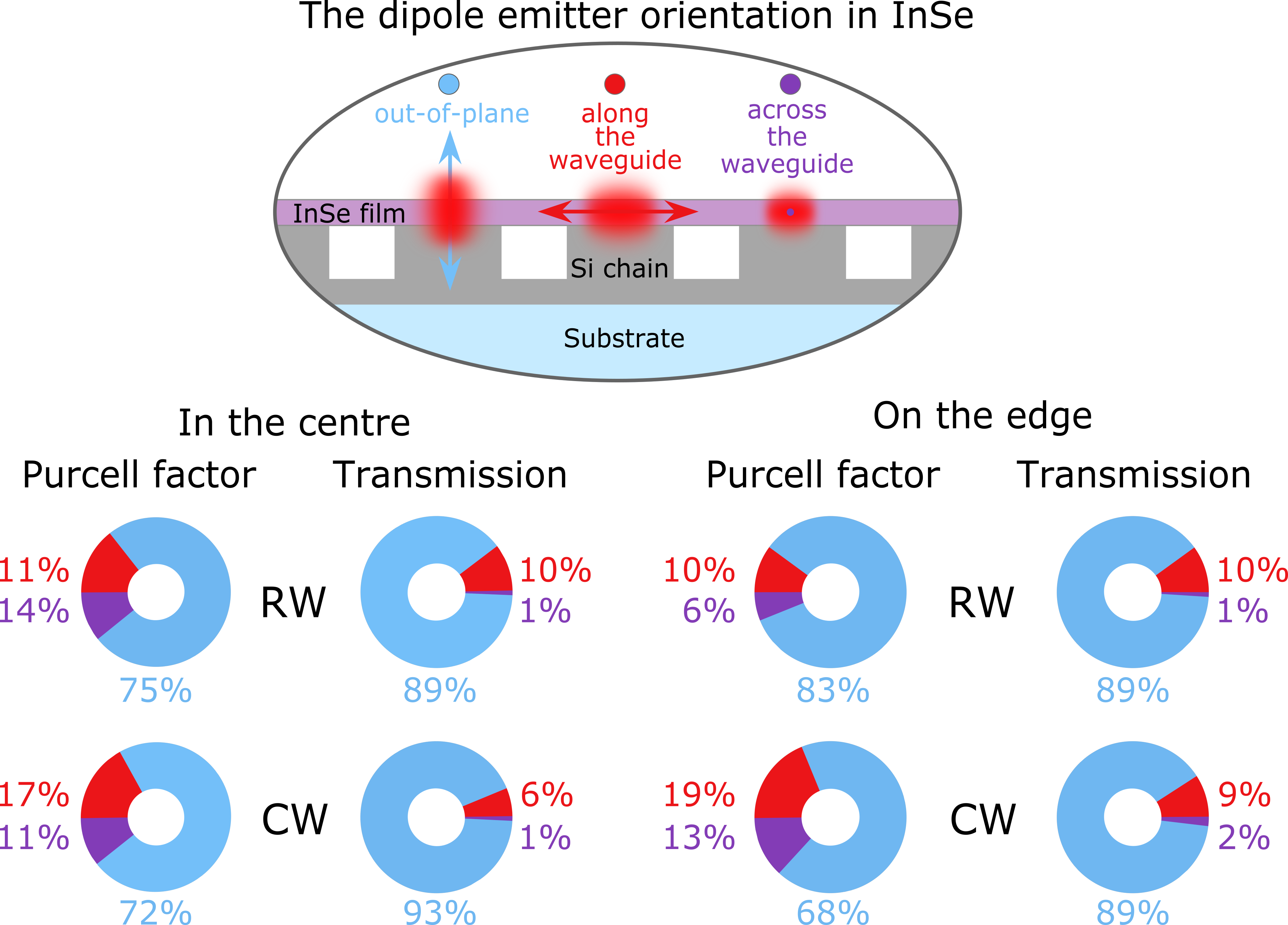} 
\caption{The influence of dipole emitter orientation in InSe on the Purcell factor and the transmission (amount of power transmitted through the waveguide, normalized to the dipole power) of the RW and the CW.}
\label{dipole_orientations}
\end{center}
\end{figure}

In order to clarify the dominant source of the PL radiation coupled into the waveguide we perform numerical modeling for variously oriented dipole emitters in InSe. In each simulation an electric dipole of the same power is placed above the central disk of the waveguide system: on the axis of symmetry and on the edge. Three types of dipole orientation relative to the waveguide system are schematically shown at the top of the Fig~{\ref{dipole_orientations}}: out-of-plane (marked with the blue color), in-plane along the waveguide (marked with the red color), and in-plane across the waveguide (marked with the violet color). 

\par\null\selectlanguage{english}
\par\null

Fig~{\ref{dipole_orientations}} shows the comparative diagrams of three types of dipole emitter orientation contributions in InSe on the Purcell factor and the transmission of RW and CW. The circle charts are the proportions of the contributions of differently oriented dipoles to the overall result. For example, in the case of a dipole placed at the center of the RW, the Purcell factor of an out-of-plane orientation is three times greater than for both in-plane orientations. The percentage of out-of-plane orientation is the highest in all cases. The Purcell factor is larger than 75$\%$ for RW and 68$\%$ for CW, while transmission exceeds 89$\%$  for RW and CW.

\section*{Discussion}

A promising development of this work is the decreasing of absorption losses in the waveguide. It can be achieved through the use of other materials, for example, silicon nitride (Si$_3$N$_4$), which is transparent at the visible and NIR wavelengths~\textsuperscript{\hyperref[csl:49]{49}}. However, in this case the design of the waveguide system will likely require modifications, since Si$_3$N$_4$ ($n\sim 2.0$) does not have as high refractive index contrast with the SiO$_2$ substrate ($n\sim 1.5$) as Si ($n\sim 3.6$). In addition, the transition to a single-photon operating mode of an excitonic emitter may be relevant for quantum optics problems. For this purpose, 2D TMDs films can be used, which can compete with more popular single-photon emitters, such as colour centres in diamonds and InAs/GaAs quantum dots~\textsuperscript{\hyperref[csl:40]{40,}}~\textsuperscript{\hyperref[csl:50]{50}}.

\section*{Conclusion}

We design and experimentally validate a resonant silicon waveguide with a central part in the form of a chain of Mie-resonant nanodisks that provides efficient optical coupling with gray excitons in InSe thin film placed on top. The disk parameters are chosen to increase the efficiency of optical coupling between dipole emitters in InSe film and resonant waveguides. Experimental results demonstrate an increase in optical coupling by a factor of~2.5 compared to a conventional rib waveguide. The result obtained in this study may be useful for a variety of integrated photonic applications.

\section*{Acknowledgements}

Authors thank the group of Isabelle Staude for help with sample fabrication. The work of M. R. was supported by a grant from the Russian Science Foundation (no. 24-72-00148, https://rscf.ru/project/24-72-00148/). The study was conducted under the state assignment of Lomonosov Moscow State University.

\section*{Conflict of interest}

Authors state no conflict of interest.

\selectlanguage{english}
\FloatBarrier
\section*{References}\sloppy
\label{csl:1}1 Kim I, et al. {Nanophotonics for light detection and ranging technology}. \textit{Nat. Nanotechnol.} 2021; \textbf{16}: 508-524.

\phantomsection
\label{csl:2}2 Estevez M, et al. {Integrated optical devices for
lab-on-a-chip biosensing applications}. \textit{Laser Photonics Rev.} 2012; \textbf{6}: 463-487.

\phantomsection
\label{csl:3}3 Karabchevsky A, et al. {On-chip
nanophotonics and future challenges}. \textit{Nanophotonics} 2020; \textbf{9}: 3733-3753. 

\phantomsection
\label{csl:4}4 Tonndorf P, et al. {On-chip
waveguide coupling of a layered semiconductor single-photon source}. \textit{Nano Lett.} 2017; \textbf{17}: 5446-5451. 

\phantomsection
\label{csl:5}5 Dionne J, et al. {Silicon-based plasmonics for on-chip photonics}. \textit{IEEE J. Sel. Top. Quantum Electron.} 2010; \textbf{16}: 295-306. 

\phantomsection
\label{csl:6}6 Wang J, et al. {On-chip silicon photonic signaling and processing: a review}. \textit{Sci. Bull.} 2018; \textbf{63}: 1267-1310. 

\phantomsection
\label{csl:7}7 Wang J, et al. {On-chip light sources for silicon photonics}. \textit{Light: Sci. Appl.} 2015; \textbf{4}: e358-e358. 

\phantomsection
\label{csl:8}8 Lieberman K, et al. {A light source smaller than the optical wavelength}. \textit{Science} 1990; \textbf{247}: 59-61. 

\phantomsection
\label{csl:9}9 Shang J, et al. {Light sources and photodetectors enabled by 2D semiconductors}. \textit{Small Methods} 2018; \textbf{2}: 1800019. 

\phantomsection
\label{csl:10}10 Koch S, et al. {Semiconductor excitons in new light}. \textit{Nat.~Mater.} 2006; \textbf{5}: 523-531. 

\phantomsection
\label{csl:11}11 Xiao J, et al. {Excitons in atomically thin 2D semiconductors and their applications}. \textit{Nanophotonics} 2017; \textbf{6}: 1309-1328. 

\phantomsection
\label{csl:12}12 Chernikov A, et al. {High-mobility three-atom-thick semiconducting films with wafer-scale homogeneity}. \textit{Phys. Rev. Lett.} 2014; \textbf{113}: 076802. 

\phantomsection
\label{csl:13}13 Xia F, et al. {Two-dimensional material nanophotonics}. \textit{Nature Photon.} 2014; \textbf{8}: 899-907. 

\phantomsection
\label{csl:14}14 Bonaccorso F, et al. {Graphene photonics and optoelectronic}. \textit{Nature Photon.} 2010; \textbf{4}: 611-622. 

\phantomsection
\label{csl:15}15 Geim A, et al. {Van der Waals heterostructures}. \textit{Nature} 2013; \textbf{499}: 419-425. 

\phantomsection
\label{csl:16}16 Rivera P, et al. {Observation of long-lived interlayer excitons in monolayer MoSe$_2$-WSe$_2$ heterostructures}. \textit{Nat. Commun.} 2015; \textbf{6}: 6242-6247. 

\phantomsection
\label{csl:17}17 Popkova A, et al. {Nonlinear Exciton-Mie Coupling in Transition Metal Dichalcogenide Nanoresonators}. \textit{Laser Photonics Rev.} 2022; \textbf{16}: 2100604. 

\phantomsection
\label{csl:18}18 Goki E, et al. {Two-Dimensional Crystals: Managing Light for Optoelectronics}. \textit{ACS Nano} 2013; \textbf{7}: 5660-5665. 

\phantomsection
\label{csl:19}19 Mak K, et al. {Photonics and optoelectronics of 2D semiconductor transition metal dichalcogenides}. \textit{Nat. Photonics} 2016; \textbf{10}: 216-226. 

\phantomsection
\label{csl:20}20 Nazarenko A, et al. {Cryogenic nonlinear microscopy of high-Q
metasurfaces coupled with transition metal dichalcogenide monolayers}. \textit{Nanophotonics} 2024; \textbf{13}: 3429-3436. 

\phantomsection
\label{csl:21}21 Peyskens F, et al. {Integration of single photon emitters in 2D layered materials with a silicon nitride photonic chip}. \textit{Nat. Commun.} 2019; \textbf{10}: 4435. 

\phantomsection
\label{csl:22}22 Wang G, et al. {In-plane propagation of light in transition metal dichalcogenide monolayers: optical selection rules}. \textit{Phys.~Rev.~Lett.} 2017; \textbf{119}: 047401. 

\phantomsection
\label{csl:23}23 Zhou Y, et al. {Probing dark excitons in atomically thin semiconductors via near-field coupling to surface plasmon polaritons}. \textit{Nat.~Nanotechnol.} 2017; \textbf{12}: 856-860. 

\phantomsection
\label{csl:24}24 Zhang X, et al. {Magnetic brightening and control of dark excitons in monolayer WSe$_2$}. \textit{Nat.~Nanotechnol.} 2017; \textbf{12}: 883-888. 

\phantomsection
\label{csl:25}25 Brotons-Gisbert M, et al. {Out-of-plane orientation of luminescent excitons in two-dimensional indium selenide}. \textit{Nat. Commun.} 2019; \textbf{10}: 3913. 

\phantomsection
\label{csl:26}26 Mudd G, et al. {Tuning the bandgap of exfoliated InSe nanosheets by quantum confinement}. \textit{Adv. Mater.} 2013; \textbf{25}: 5714-5718. 

\phantomsection
\label{csl:27}27 Segura A, et al. {Electron scattering mechanisms in n-type indium selenide}. \textit{Phys. Rev. B} 1984; \textbf{29}: 5708.

\phantomsection
\label{csl:28}28 Kuroda N, et al. {Resonance Raman scattering study on exciton and polaron anisotropies in InSe}. \textit{Solid State Commun.} 1980; \textbf{34}: 481-484. 

\phantomsection
\label{csl:29}29 Kress-Rogers E, et al. {Cyclotron resonance studies on bulk and two-dimensional conduction electrons in InSe}. \textit{Solid State Commun.} 1982; \textbf{44}: 379-383. 

\phantomsection
\label{csl:30}30 Leisgang N, et al. {Optical second harmonic generation in encapsulated single-layer InSe}. \textit{AIP Advances} 2018; \textbf{8}: 105120. 

\phantomsection
\label{csl:31}31 Tamalampudi S, et al. {High Performance and Bendable Few-Layered InSe Photodetectors with Broad Spectral Response}. \textit{Nano Lett.} 2014; \textbf{14}: 2800-2806. 

\phantomsection
\label{csl:32}32 Lei S, et al. {Evolution of the Electronic Band Structure and Efficient Photo-Detection in Atomic Layers of InSe}. \textit{ACS Nano} 2014; \textbf{8}: 1263-1272.

\phantomsection
\label{csl:33}33 Feng W, et al. {Ultrahigh photo-responsivity and detectivity in multilayer InSe nanosheets phototransistors with broadband response}. \textit{J. Mater. Chem. C} 2015; \textbf{3}: 7022-7028.

\phantomsection
\label{csl:34}34 Gulkin D, et al. {Mie-driven directional nanocoupler for Bloch surface wave photonic platform}. \textit{Nanophotonics} 2021; \textbf{10}: 2939-2947.

\phantomsection
\label{csl:35}35 Obydennov D, et al. {Asymmetric Silicon Dimers Made by Single-Shot Laser-Induced Transfer Demultiplex Light of Different Wavelengths}. \textit{Adv.~Optical~Mater.} 2023; \textbf{12}: 2302276.

\phantomsection
\label{csl:36}36 Kuznetsov A, et al. {Optically resonant dielectric nanostructures}. \textit{Science} 2016; \textbf{354}: aag2472.

\phantomsection
\label{csl:37}37 Krasnok A, et al. {All-dielectric optical nanoantenna}. \textit{Opt. Express} 2012; \textbf{20}: 20599-20604.

\phantomsection
\label{csl:38}38 Sergaeva O, et al. {Resonant dielectric waveguide-based nanostructure for efficient interaction with color centers in nanodiamonds}. \textit{Phys. Chem. Math.} 2019; \textbf{10}: 266.

\phantomsection
\label{csl:39}39 Bakker R, et al. {Resonant Light Guiding Along a Chain of Silicon Nanoparticles}. \textit{Nano Lett.} 2017; \textbf{17}: 3458-3464.

\phantomsection
\label{csl:40}40 Gartman A, et al. {Efficient Integration of Single-Photon Emitters in Thin InSe Films into Resonance Silicon Waveguides}. \textit{JETP Lett.} 2020; \textbf{112}: 693-698

\phantomsection
\label{csl:41}41 Gartman A, et al. {Efficient Light Coupling and Purcell Effect Enhancement for Interlayer Exciton Emitters in 2D Heterostructures Combined with SiN Nanoparticles}. \textit{Nanomaterials} 2023; \textbf{13}: 1821

\phantomsection
\label{csl:42}42 Song C, et al. {The optical properties of few-layer InSe}. \textit{J. Appl. Phys.} 2020; \textbf{128}: 060901. 

\phantomsection
\label{csl:43}43 Zolyomi V, et al. {Electrons and phonons in single layers of hexagonal indium chalcogenides from ab initio calculations}. \textit{Phys. Rev. B} 2014; \textbf{89}: 205416. 

\phantomsection
\label{csl:44}44 Borodin B, et al. {Indirect-to-direct band-gap transition in few-layer $\ensuremath{\beta}$-InSe as probed by photoluminescence spectroscopy}. \textit{Phys. Rev. Mater.} 2024; \textbf{8}: 014001. 

\phantomsection
\label{csl:45}45 Wang Y, et al. {Interference enhancement of Raman signal of graphene}. \textit{Appl. Phys. Lett.} 2008; \textbf{92}: 043121. 

\phantomsection
\label{csl:46}46 Song C, et al. {Drastic enhancement of the Raman intensity in few-layer InSe by uniaxial strain}. \textit{Phys. Rev. B} 2019; \textbf{99}: 195414. 

\phantomsection
\label{csl:47}47 Yao P, et al. {On-chip single photon sources using planar photonic crystals and single quantum dots}. \textit{Laser Photon. Rev.} 2010; \textbf{4}: 499. 

\phantomsection
\label{csl:48}48 Chakraborty C, et al. {Advances in quantum light emission from 2D materials}. \textit{Nanophotonics} 2019; \textbf{8}: 2017-2032. 

\phantomsection
\label{csl:49}49 Subramanian A, et al. {Low-Loss Singlemode PECVD Silicon Nitride Photonic Wire Waveguides for 532-900 nm Wavelength Window Fabricated Within a CMOS Pilot Line}. \textit{IEEE Photonics J.} 2013; \textbf{5}: 2202809-2202809. 

\phantomsection
\label{csl:50}50 Aharonovich I, et al. {Solid-state single-photon emitters}. \textit{Nat. Photonics} 2016; \textbf{10}: 631-641. 

\end{bibunit}

\end{document}